\documentclass[a4paper,fleqn,usenatbib]{mnras}

\usepackage[T1]{fontenc}
\usepackage{ae,aecompl}

\usepackage{graphicx}	
\usepackage{amsmath}	
\usepackage{amssymb}	
\usepackage{comment}
\usepackage{color}

\title[]{A common origin for the circumnuclear disc and the nearby molecular clouds in the Galactic Centre}

\author[A. Ballone et al.]{Alessandro Ballone$^{1,2}$, Michela Mapelli$^{1,2,3,4}$, Alessandro Alberto Trani$^{5}$
\\
$^{1}$INAF - Osservatorio Astronomico di Padova, Vicolo dell'Osservatorio 5, I-35122 Padova, Italy\\
$^{2}$INFN - Padova, Via Marzolo 8, I--35131 Padova, Italy\\
$^{3}$Physics and Astronomy Department Galileo Galilei, University of Padova, Vicolo dell'Osservatorio 3, I--35122, Padova, Italy\\
$^{4}$Institute f\"ur Astro- und Teilchen Physik, Universit\"at Innsbruck, Technikerstrasse 25/8, A-6020 Innsbruck, Austria\\
$^{5}$Department of Astronomy, Graduate School of Science, The University of Tokyo, 7-3-1 Hongo, Bunkyo-ku, Tokyo, 113-0033,
Japan\\}

\date{Accepted XXX. Received YYY; in original form ZZZ}

\pubyear{2018}

\begin{document}
\label{firstpage}
\pagerange{\pageref{firstpage}--\pageref{lastpage}}
\maketitle

\begin{abstract}
The origin of the molecular clouds orbiting SgrA* is one of the most debated questions about our Galactic Centre. Here, we present the hydrodynamic simulation of a molecular cloud infalling towards SgrA*, performed with the adaptive-mesh-refinement code RAMSES. Through such simulation, we propose that the circumnuclear disc and the +20 km/s cloud originated from the same tidal disruption episode, occurred less than 1 Myr ago. We also show that recent star formation is to be expected in the +20 km/s cloud, as also suggested by recent observations.
\end{abstract}




\section{Introduction}\label{intro}

The inner few parsecs of the Galactic Centre are an extremely dense environment, hosting a supermassive black hole (SMBH) with mass of
around $4\times 10^6 \; M_{\odot}$ \citep{Boehle16,Gillessen17}, often referred to as SgrA* (i.e., its radio point source counterpart), a large number of late-type stars, few hundreds young and massive stars
with very peculiar properties and a widely variegated interstellar medium \citep[for general reviews, see][]{Morris96, Genzel10,Mapelli16b}.

The late-type stars are in an almost isotropic \citep[though some triaxality is observed in the inner 2 pc;][]{FeldmeierKrause17} dense nuclear star cluster, with mass of around $2\times 10^7 \; M_{\odot}$ in the inner 10 pc \citep{Schodel07,Schodel09,Schodel14,Feldmeier14,Chatzopoulos15,Fritz16,FeldmeierKrause17,GallegoCano18}.

On the other hand, most young stars reside in the inner pc and a considerable percentage (20-50\%) of these lie in one disc  \citep{Bartko09,Yelda14}. Most young stars are classified as O-type (many even in a Wolf-Rayet phase) and have an age of
few Myr (4-8 Myr: \citealp{Paumard06}; 2.5-6 Myr: \citealp{Lu13}), hence suggesting that their birth
occurred in a recent, single star formation episode. In addition to the disc, a second group of few tens
of B-type stars \citep[with age $\lesssim$ 20 Myr;][]{Habibi17}, the so-called S-stars, reside in the
inner 0.05 pc of the Galaxy and have isotropically oriented, highly eccentric orbits around the SMBH
\citep{Ghez08,Gillessen09,Gillessen17}.

These young stars are surrounded by a circumnuclear disc (CND) of molecular and neutral gas and dust, sitting at 1.5-4 pc from SgrA* \citep[e.g.,][]{Gatley86, Guesten87, Jackson93, Herrnstein02,Shukla04, Christopher05, Montero09,Oka11,Martin12, Lau13,Mills13}. The CND also contains several tens of $M_{\odot}$ of ionized gas in clumpy filaments/arms observable in radio and infrared continuum and emission lines \citep[i.e., the so-called ``minispiral''; see, e.g.,][]{Lo83,Serabyn85, Paumard04, Zhao09,Zhao10,Lau13,Tsuboi16,Moser17}.

Within the inner 20 pc in projected distance from SgrA*, two large molecular clouds, with mass $\gtrsim 10^5 \; M_{\odot}$ are also observed, often called +50 km/s (or M-0.02-0.07) and +20 km/s (or M-0.13-0.08) \citep{Gusten80, Gusten83, Liszt85,Sandqvist89,Lis94,Novak00}. In particular, the +50 km/s lies at a distance of roughly 7 pc from SgrA*, it has a roughly spherical shape and a radius of around 5 pc, though a large portion of the cloud seems to be compressed by the radio non-thermal shell SgrA East \citep[interpreted as a supernova remnant; e.g.,][]{Ho85, Genzel90, Serabyn92,McGary01, Lee03, Lee08}. The +20 km/s cloud lies around 12 pc away from SgrA* and it is elongated ($\approx$ 15 pc $\times$ 7.5 pc) in the direction of the SMBH.
Recent observations also suggest the presence of few protostars embedded in the +20 km/s cloud \citep{Lu15,Lu17}.
According to some authors, these two clouds orbit around the Galactic Centre at distances of 50-100 pc from the centre of the Galaxy, possibly in a larger belt composed of many molecular clouds \citep{Gusten80, Gusten81, Zylka90, Kruijssen15,Henshaw16, Kruijssen19}. On the other hand, many authors agree on rather placing them at distances $\lesssim 20$ pc, based on their possible interaction with SgrA East \citep[e.g.,][]{Genzel90, McGary01,Herrnstein05, Lee08} and on their internal velocity pattern and the presence of several streamers of gas connecting them to the CND \citep[e.g.,][]{Okumura89, Coil99,Coil00,Ferriere12, Oka12,Liu12,Hsieh17, Takekawa17,Tsuboi18}. In particular, the so-called southern streamer seems to continously extend from the +20 km/s cloud to the CND, feeding it and possibly the inner pc of the Galaxy \citep[in particular,][]{Okumura91, Coil99,Coil00,Montero09}.

A connection between the inner few parsecs of the galaxy and the close-by molecular clouds is also supported by pure theory. Molecular clouds ``engulfing'' the SMBH are often invoked to explain the formation of the disc of young stars \citep{Sanders98,Wardle08,Bonnell08,Hobbs09,Alig11,Mapelli12,Lucas13,Alig13} and/or the origin of the CND (\citealp{Sanders98,Wardle08,Mapelli16c,Trani18}; however, other authors modeled the CND as a long-lived stable structure, ``carved'' by feedback from the young stars or by tidal shear, see, e.g., \citealp{Vollmer01,Vollmer02, Blank16}).

Here we draw on this ideas to show that the CND and the +20 km/s cloud might be explained by the same tidal distruption episode.

\section{Methods}

The simulation we used for this study is part of a set of runs of different molecular clouds in the Galactic Centre, with different masses, radii and initial orbital conditions that were performed to investigate the impact of these on star formation in the inner (tens of) parsecs of our Galaxy (Ballone et al., in preparation). Among this set we found this particular run that seems to qualitatively match some of the properties of the observed CND and +20 km/s cloud. We must then stress that the parameteres of this simulation were not fine-tuned and not all the observed features will be exactly reproduced.

\subsection{Numerical setup}\label{numset}

The simulation has been run with the adpative mesh refinement (AMR) code RAMSES \citep{Teyssier02}. The domain is cubic, with a size of 100 pc$^3$, and it is centred on the position of the SMBH. The resolution of the grid spans 9 levels of refinement, from level 6, corresponding to a resolution of $\Delta x_6=100/2^6\approx 1.56$ pc, to level 14, corresponding to a resolution of $\Delta x_{14}=100/2^{14}\approx 6.1\times 10^{-3}$ pc. In order to get a good refinement of the dense (and possibly collapsing) material, without getting too high computational cost, we used a so-called ``quasi-Lagrangian'' scheme \citep[e.g.,]{Teyssier15}, that forces any cell to be refined if the enclosed mass is higher than a certain value $m_l$. For level $l=[6,14]$, we chose $m_l=k_l m_{ref}$, with $m_{ref}\approx 5.82\times 10^{-2} \; M_{\odot}$ and $k_l=\{1,1,1,10^4,10^2,10,3,1\}$. Such choice of $k_l$ has been based on ensuring that the molecular cloud is refined, at the beginning, at least to level 8; at the same time, it guarantees that only the densest parts of the cloud, after turbulence develops (see Section \ref{physet}), are followed at the highest resolution\footnote{$m_{ref}$ is roughly 1/64 of the jeans mass of the smallest cells, at level 14.}. We adopted a Lax-Friedrichs Riemann solver, which is the most diffusive, but also the only stable one for our problem, and a MinMod slope limiter for the piecewise linear reconstruction of the Godunov scheme. 

\begin{figure}{}
\begin{center}
\includegraphics[scale=0.56]{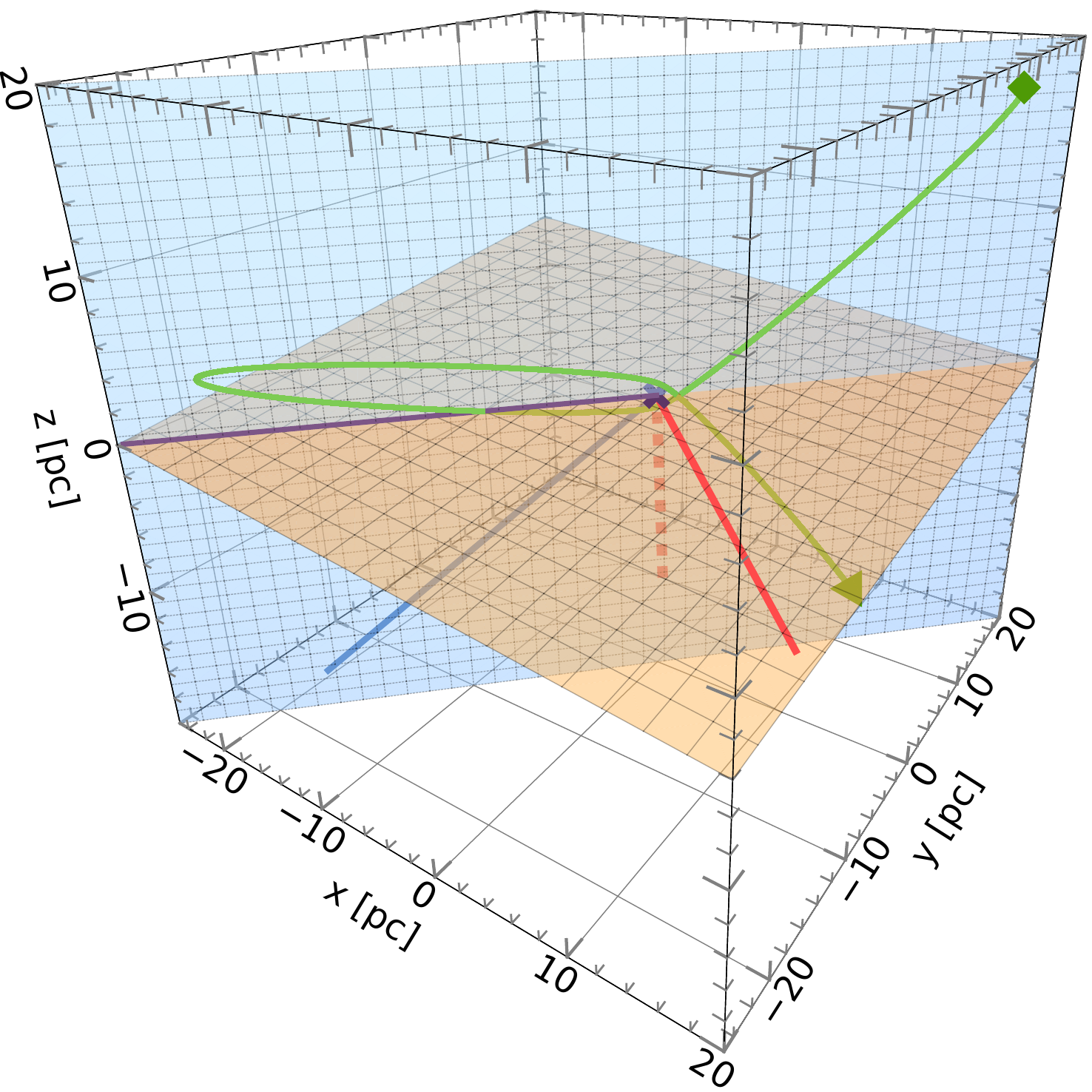}
\caption{Plot of the orbital orientation in the computational domain. The green curve marks the orbit of the initial center of the cloud, from the beginning of the simulation (green square) up to 0.8 Myr (reached at the tip of the arrow). The orbital plane is marked in light blue, with the blue solid line marking the direction opposite to the initial apocentre (we remind the reader that the orbit is precessing). The orange plane is the plane of the sky in our projection (see section \ref{results}), the red solid line marks the direction of the Galactic latitude (longitude equal to 0) axis and the red dotted line marks the direction of the line-of-sight. The purple solid line marks the direction of the ascending node and the purple cross marks the position of the SMBH (and the center of the computational domain).
}\label{orbitproj}
\end{center}
\end{figure}

\begin{figure*}
\begin{center}
\includegraphics[scale=0.6]{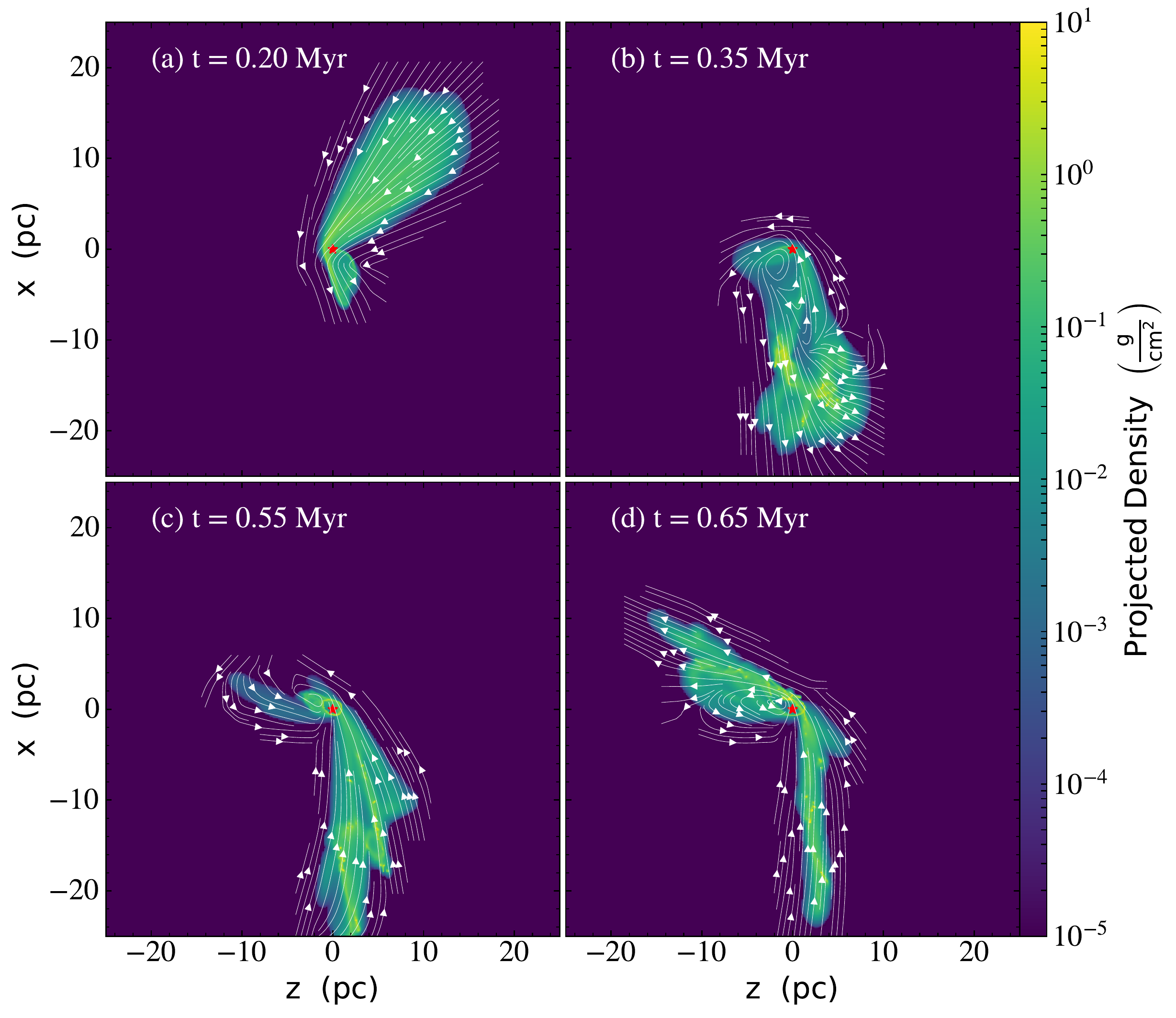}
\caption{Time evolution of the cloud plunge. The different panels show a density projection along the $y$ axis of the computational domain, at $t= 0.15$ (panel a), 0.20 (panel b), 0.35 (panel c) and 0.55 (panel d) Myr. The red star marks the position of the SMBH.
}\label{timeev}
\end{center}
\end{figure*}

\subsection{Physical setup}\label{physet}

The gravity of the SMBH has been simulated by introducing a particle of mass $M_{SMBH}=4.3\times 10^6 \; M_{\odot}$ \citep[this is roughly the estimate by][]{Gillessen09,Gillessen17} from the beginning of the simulation. Such particle is a ``sink'' (hence it can accrete matter, see later) and has been anchored to the centre ($x=y=z=0$) of the simulation domain, to avoid artificial kicks due to pure numerical errors.
The gravitational contribution of the stellar cusp of stars and the nuclear stellar disc \citep[i.e., the so-called ``nuclear bulge'';][]{Launhardt02} has also been included as a spherically symmetric rigid potential. The adopted density profile for both components is $\rho_{NB}(r)= 2.8 \times 10^6 (r/r_{NB})^{-\gamma} \; M_{\odot} \;\mathrm{pc}^{-3}$ (where $r$ is the distance from the centre of the computational domain), with $r_{NB}=0.22$ pc, $\gamma=1.2$ for $r < r_{NB}$ and $\gamma=1.75$ for $r > r_{NB}$ \citep[e.g.,][]{Schodel07}. 
For completeness, we also included the Galactic bulge potential, as a Hernquist sphere with density profile $\rho_{GB}(r)=2.9 \times 10^{10} r_{GB}/[2\pi r(r+r_{GB})^3] \;M_{\odot}$, where $r_{GB}=0.7$ kpc \citep{Mapelli16c}.

The simulated molecular cloud has mass $M_{cl}=10^5\; M_{\odot}$ and radius $R_{cl}=10$ pc (corresponding to a density $\rho_{cl}\approx 1.6\times 10^{-21} \; \mathrm{g \; cm^{-3}}$) and it has been initially placed at a distance $r_{cl}=30$ pc ($x_{cl}=y_{cl}=z_{cl}=30/\sqrt{3}$ pc) from the SMBH, with tangential velocity $v_{t,cl}=10 \mathrm{\; km \; s^{-1}}$ ($v_{x,cl}=v_{y,cl}=10/\sqrt{6} \mathrm{\; km \; s^{-1}}$, $v_{z,cl}=-20/\sqrt{6} \mathrm{\; km \; s^{-1}}$). This means that the cloud is initially on one of the diagonals of our computational cubic domain and that the orbit lies on the x=y plane (see Fig. \ref{orbitproj}). The initial pericenter distance of the cloud center of mass is at a distance of  $\approx 0.76$ pc from SgrA*. 
We gave the cloud a temperature of $T_{cl}=100$ K. The cloud is initially given a random Gaussian, divergence-free turbulent velocity field, with power spectrum $P(k)=||\delta{}^2_v  || \propto{} k^{-4}$. This is the so-called Burgers power spectrum \citep{Burgers48} and it is the one that seems to reproduce, in molecular clouds, the observed trend of the velocity dispersion with the cloud size and the size of its subregions \citep[e.g.,][]{Larson81}. Such turbulent field is needed to produce density peaks in the cloud that later collapse by self-gravity and lead to star formation all over the cloud. At the beginning of the simulation, the cloud is marginally bound (i.e., its turbulent kinetic energy is equal to its potential energy). This also means that the initial turbulent kinetic energy is a factor of 50 larger than the thermal energy of the cloud.

Though the Galactic Centre and its molecular gas are certainly embedded in a complex gaseous environment, for the sake of simplicity we just included in the simulation the molecular gas of the initial cloud, surrounding it with a low-density ``empty atmosphere'', with density $\rho_{atm}=1.6\times 10^{-27} \; \mathrm{g \; cm^{-3}}$ and temperature $T_{atm}=10^6$ K.

To model the thermodynamics of the simulation we decided to adopt a simple isothermal equation of state (EoS). This assumption is not capturing the complex heating-cooling interplay affecting the temperature of molecular gas; however, most of the clouds in the Galactic Centre show a temperature of $\approx 50-150$ K \citep{Ao13}, which is probably the temperature of thermal equilibrium \citep[e.g.,][]{Larson85,Koyama00} in this region.

Star formation is treated in RAMSES through the sink particle algorithm by \citet{Bleuler14}. This algorithm is quite sophisticated. In extreme synthesis, it checks for cells in the computational domain with density higher than $\rho_{sink}$, groups these cells into clumps and performs for these a ``collapse'', a ``proximity'' and a ``virial''\footnote{The ``virial'' check, as performed in RAMSES' sink particle algorithm, is particularly useful, since it also takes into account the tidal force from the background potential, which is particularly important for our problem.} check to ensure that the clump will collapse onto a point-like object. For our simulation we adopted $\rho_{sink}=3\times 10^{-17} \; \mathrm{g \; cm^{-3}}$, which is approximately the density for which the Jeans length equals 4 times the minimum adopted cell size ($\Delta x_{14}$, see Section \ref{numset}). We performed few tests and found that the value of $\rho_{sink}$ does not significantly impact the result on star formation. The sinks forming during the simulation are all born with an initial ``seed'' mass of $3.5 \; M_{\odot}$, which is about the jeans mass at the highest resolution. During the evolution, sink particles can still accrete gas, according to the Bondi-Hoyle accretion algorithm implemented in RAMSES \citep[again, for details, check][]{Bleuler14}. To avoid spurious accretion due to lack of resolution, we set an upper limit to the sink mass of 150 $M_{\odot}$: if a sink reaches this upper limit during the simulation, no further accretion onto that sink is allowed. This is true for all sinks, except the SMBH, which, as already mentioned, also behaves as a sink in the simulation, hence it is gaining mass with time.

\section{Results}\label{results}

In Fig. \ref{timeev}, we show the time evolution of the simulation. The cloud gets immediately stretched by the tidal force of the background gravitational potential of the SMBH and the stars, reaching the maximal squeezing at the closest distance from the SMBH (see panel a). 
The cloud reaches its first apocentre after roughly 0.4 Myr, though the parts of the cloud with lowest angular momentum already start falling back towards the SMBH and whirl around it (panel b and c).  At this stage, the apocentre position has significantly changed, compared to the initial one, due to very fast orbital precession induced by the massive stellar cusp \citep{Gualandris12, Trani18}. A circumnuclear ring is already clearly visible after around 0.5 Myr from the beginning of the simulation, while most of the cloud is still elongated towards the SMBH and has not fallen back yet. After this time, the bulk of the cloud has reached its second pericentre passage and its leading part is already receding again, though a fraction of this post-pericentre material is streaming back towards the SMBH (panel d).

In order to try to reproduce the current configuration of the Galactic Centre, we had to define a new reference system corresponding to the actual line of sight (with versor $\hat{s}$), Galactic longitude (with versor $\hat{l}$) and latitude (with versor $\hat{b}$). In this way, we could define our orbit through three angles defining its orientation with respect to the new ``observational'' reference system. Specifically, $i$ was defined as the inclination between the orbital plane and the plane of the sky, $\Omega$ was defined as the angle, in the sky plane, between the direction of the ascending node and the Galactic latitude (l=0) axis and $\omega$ is the angle, in the orbital plane, between the direction of the ascending node and the direction opposite to the initial apocentre (the pericentre direction is changing with time, due to precession; see above). To convert physical lengths to longitude and latitude, we assumed a Galactic Centre distance of 8.32 kpc \citep{Gillessen17}.

In the following, we will present our main scenario, which is able to explain the formation of the CND and its interaction with the +20 km/s cloud, as well as to reproduce some observational properties of these two molecular structures.

\begin{figure}{}
\begin{center}
\hspace*{0.105cm}
\includegraphics[scale=0.55]{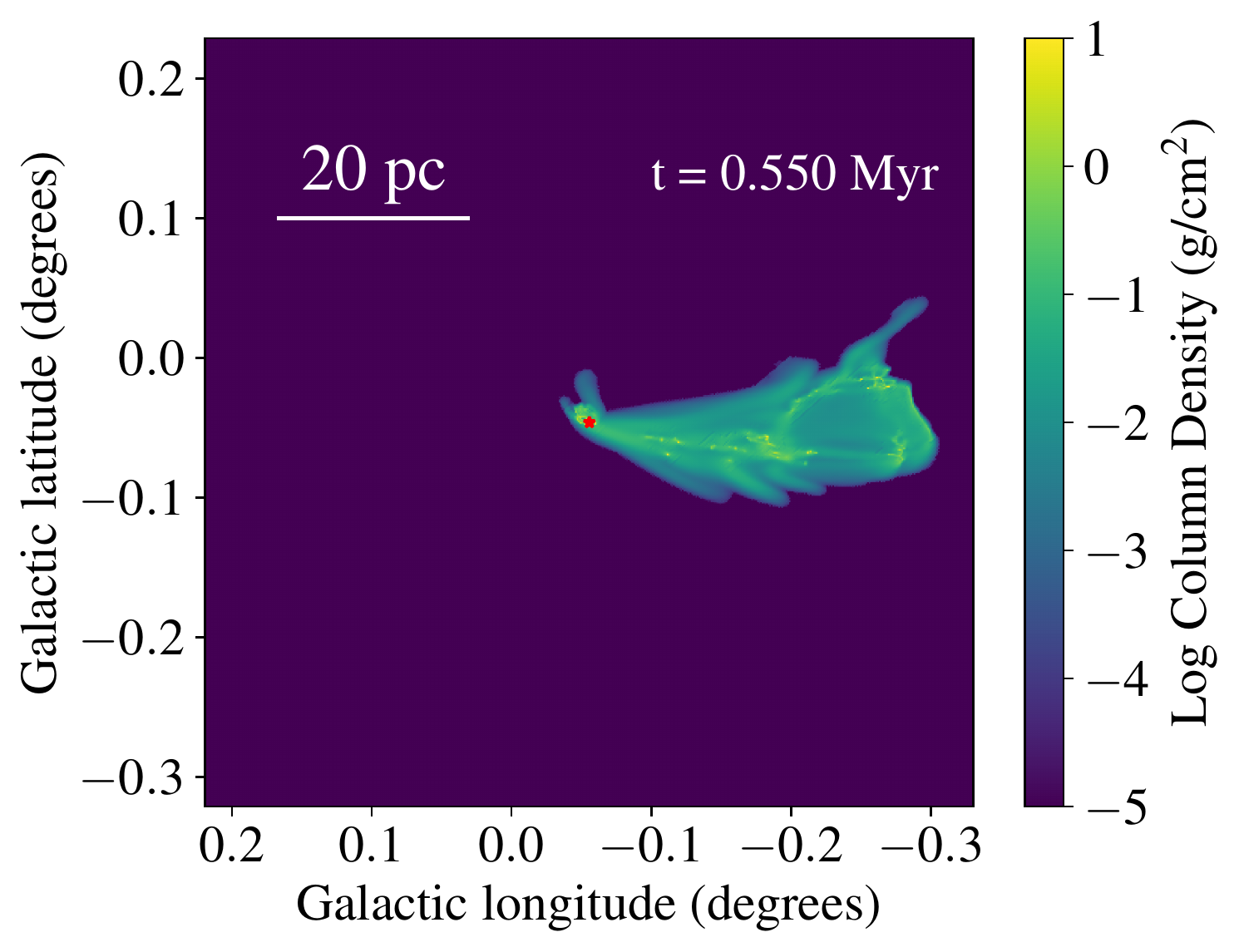}
\includegraphics[scale=0.55]{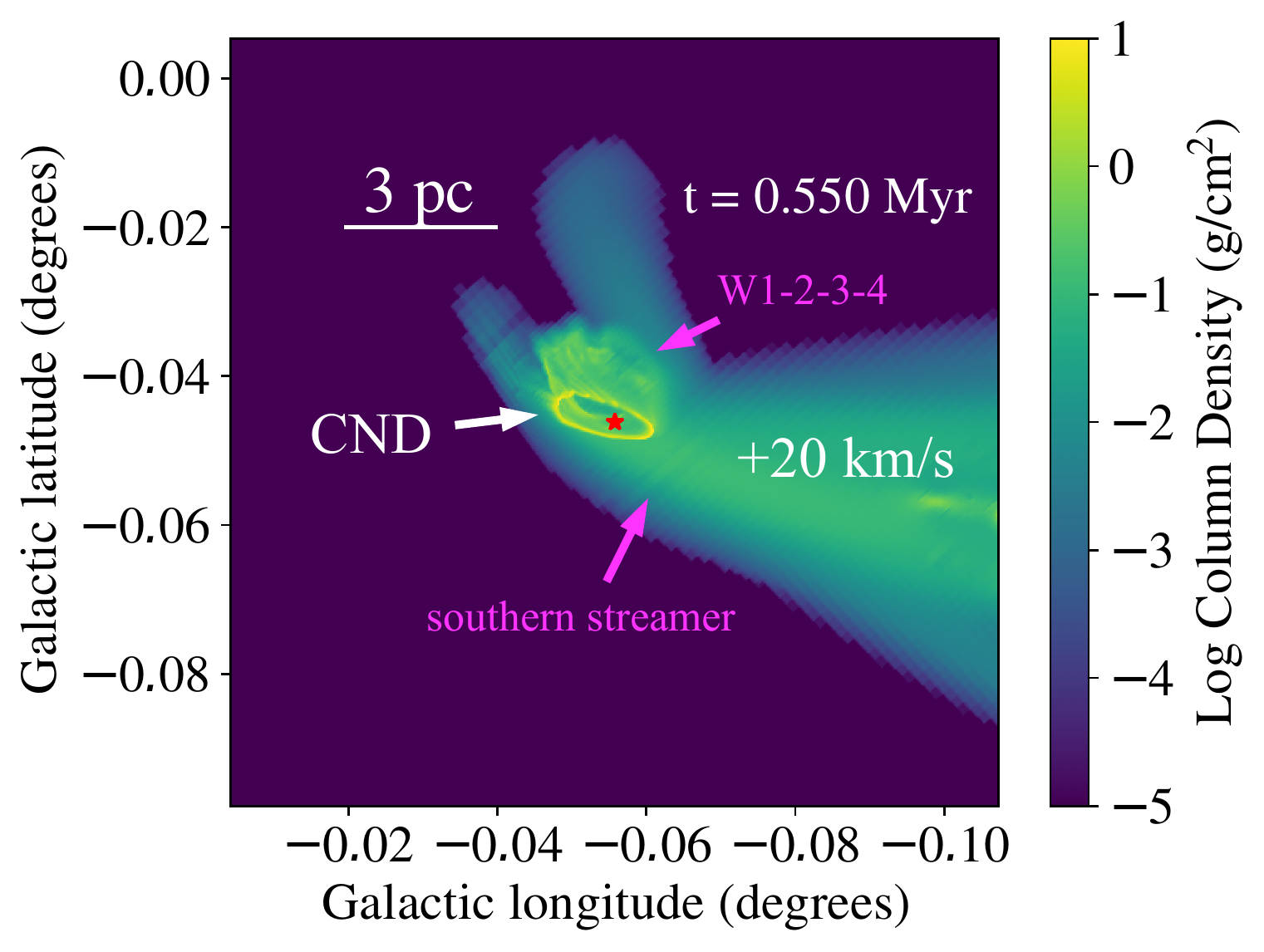}
\caption{Density projection along the line-of-sight in the new reference system (see text in section \ref{geomno50}) for our scenario, at t = 0.550 Myr. The axes represent the Galactic longitude and latitude. The upper panel shows all the gas in the computational domain, the lower panel is just a zoom-in around the inner 16 pc of the domain. The red star labels the position of the SMBH.
}\label{proj_no50}
\end{center}
\end{figure}

\subsection{Geometry}\label{geomno50}

To simultaneously reproduce the +20 km/s cloud and the CND we had to choose $(x,y,z)_{\hat{s}}\approx(-0.050,0.132,-0.990)$, $(x,y,z)_{\hat{l}}\approx(-0.661,-0.747,-0.007)$ and $(x,y,z)_{\hat{b}}\approx(0.749,-0.651,-0.125)$, hence with $\hat{b}=\hat{s}\times\hat{l}$.
This corresponds to three orbital orientation angles (see above) $i\approx 97.4^{\circ}$, $\Omega\approx 267.2^{\circ}$ and $\omega\approx 35.3^{\circ}$. Fig. \ref{orbitproj} shows the orbital plane and the sky plane, as well as few reference directions, in the adopted projection. 

This projection led to the density maps shown in Fig. \ref{proj_no50}. In this case, the molecular cloud is ``wrapping'' around the CND from the south-east as it seems to be suggested by the observations \citep[cf. with the southern streamer/ridge; e.g.,][]{Coil00, Ferriere12, Liu12, Hsieh17}. The zoomed map on the bottom panel of Fig. \ref{proj_no50} also shows several smaller streamers in the north-west part of the CND, that quite well resemble the so-called western streamers (W-1, W-2, W-3 and W-4; see Fig. 2 in \citealp{Liu12} and Fig. 4 in \citealp{Hsieh17}). The inclination of the ring with respect to the sky plane is also similar to the one of the observed CND. On the other hand, the molecular cloud is bigger than the +20 km/s. This problem might potentially be solved by slightly different initial condtions. For example, in this scenario, the +20 km/s cloud can (and does) have a maximum elongation equal to its orbital apocentre distance. Hence, reducing the apocentre distance to the actual length of the +20 km/s cloud (i.e., about 15 pc) would make the simulated cloud less elongated. Furthermore, an initially more compact cloud would be stretched less by the tidal field, also reducing the extension of the simulated cloud.

\begin{figure}
\begin{center}
\includegraphics[scale=0.55]{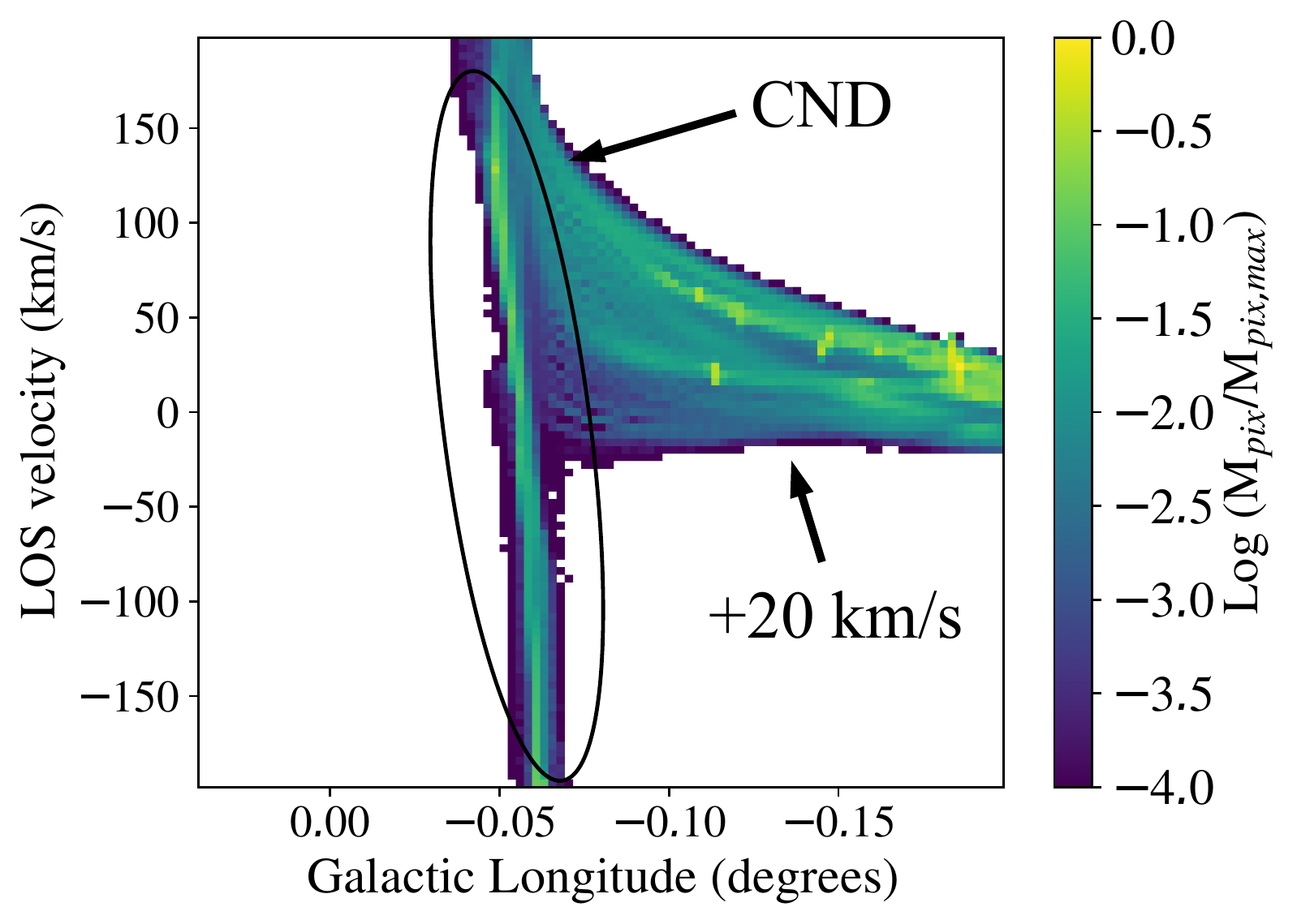}
\caption{Longitude-velocity diagram for our scenario, at t = 0.550 Myr. The colormap shows the mass in every pixel of the diagram, scaled to the maximum value. At this time our simulation is able to reproduce some of the features of the observed CND and +20 km/s cloud.
}\label{pv_no50}
\end{center}
\end{figure}

\begin{figure}
\begin{center}
\includegraphics[scale=0.55]{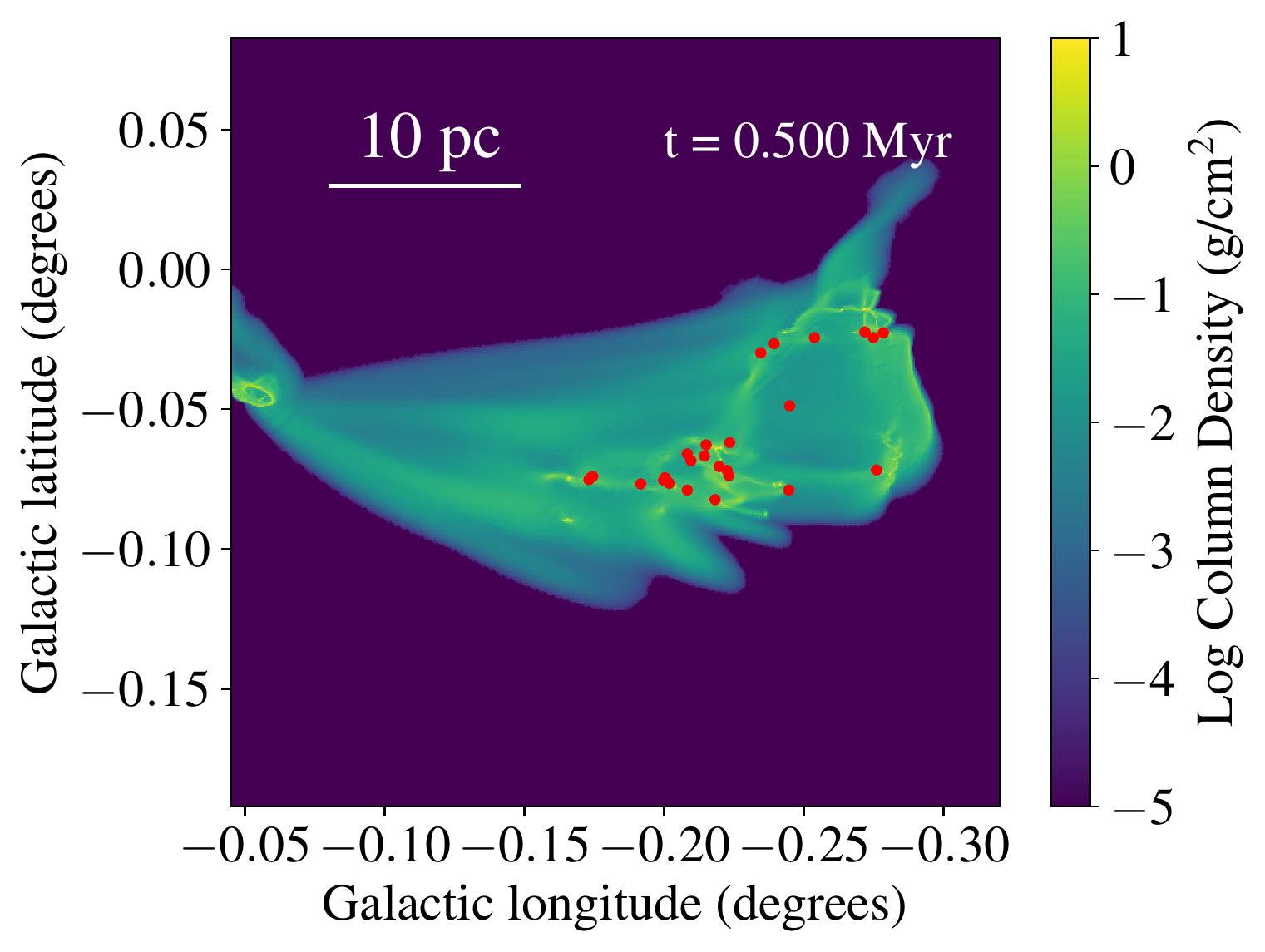}
\caption{Density projection, at t = 0.5 Myr, along the line-of-sight for the scenario described in section \ref{secno50}. This plot shows the position of the sink particles formed in the simulation (red dots). The axes represent the Galactic latitude and longitude, though in physical units (pc). 
}\label{stars_no50}
\end{center}
\end{figure} 

\subsection{Longitude-velocity diagram}

The longitude-velocity diagram in Fig. \ref{pv_no50} is produced including all the gas in our simulation. This plot clearly shows the ring in the inner 2 pc as an oblique line going from positive line-of-sight velocity (around +100 km/s) and positive latitude to negative line-of-sight velocity (around -150 km/s) and negative latitude. This is roughly comparable to what is observed for the CND (cf. with Fig. 1 in \citealp{Oka11}, Fig. 8 in \citealp{Hsieh17} and Fig. 1 in \citealp{Takekawa17}) and it is coherent with the orbital motion of the cloud (i.e., the CND is corotating with the +20 km/s cloud). The infalling molecular cloud is occupying the right part of the PV diagram, with a median line-of-sight velocity of roughly +20 km/s, though with a moderately larger spread in velocity, compared to the observations.
The longitude-velocity diagram also shows the presence of gas with higher velocity, compared to the CND. This is gas that is orbiting the SMBH inside (i.e., faster than) the ring. We know that gas inside the CND is present but at higher temperatures (i.e., the minispiral, see section \ref{intro}), most probably ionized by the disc of young massive stars. Due to our simplified assumption of isothermal gas and the lack of such ionization sources, this material is appearing in our simulated longitude-velocity diagram, but it would not if we were including such more sophisticated physical ingredients and produce different longitude-velocity diagrams for different gas phases.
We do not detect, in this diagram, any feature similar to the negative-longitude extension (NLE), as visible in the longitude-velocity diagram in Fig. 1 of \citet{Takekawa17}.

\subsection{Star formation in the +20 km/s cloud}

In Fig. \ref{stars_no50} we show the position, at t = 0.5 Myr, of all the 32 sink particles formed in the simulation (see section \ref{physet}) at this time. These sink particles should trace (proto)stars, born by the collapse of the gas in local density peaks seeded by the turbulence. All the sinks shown in this plot are younger than 0.3 Myr, i.e., they formed during or after the first pericentre passage of the cloud. This is compatible with the observational indications of ongoing star formation in the +20 km/s cloud, presented by \citet{Lu17}. In particular, these authors found 13 H$_2$O masers coincident with submm continuum detected clumps and further 5 gravitationally bound clumps detected only in the submm continuum, for a total of 18 potential protostars. Such number is roughly of the order (within a factor of 2) of the number of sinks formed in the simulation\footnote{A possible caveat is that our resolution does not allow to count single stars, i.e., our sinks might also represent very small (proto)clusters of stars. Furthermore, we lack stellar feedback, which might also play a role in this sense. Nonetheless, the sink location should capture the regions where collapse has occurred, which is more or less what has been done by \citet{Lu17}, looking at the gravitationally bound clumps.}

\begin{figure}
\begin{center}
\hspace*{0.105cm}
\includegraphics[scale=0.55]{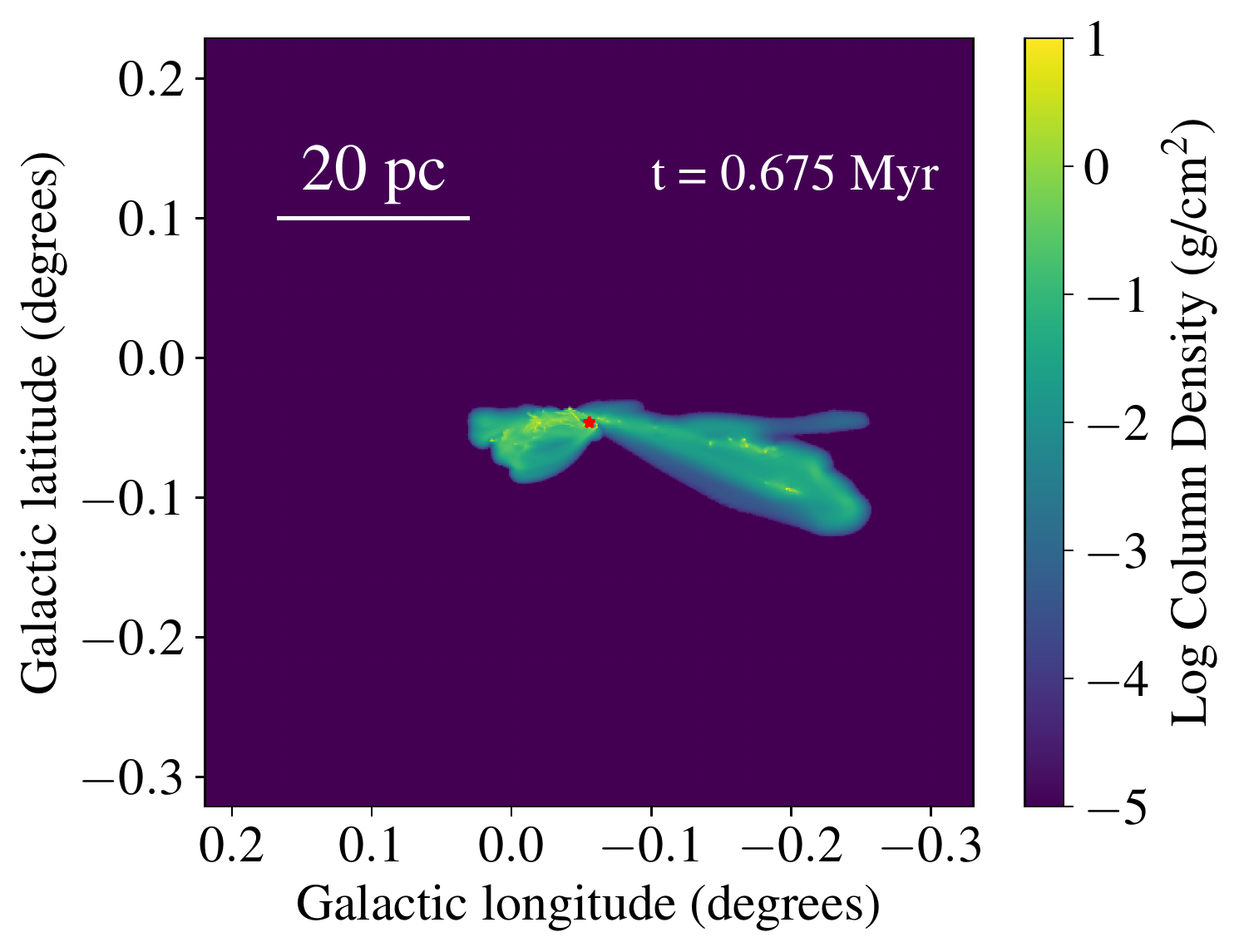}
\caption{Density projection along the line-of-sight, at t = 0.675 Myr, in a different reference system compared to our reference scenario (see Appendix \ref{app50}). The axes represent the Galactic longitude and latitude.
}\label{proj_with50}
\end{center}
\end{figure}

\begin{figure}
\begin{center}
\includegraphics[scale=0.5]{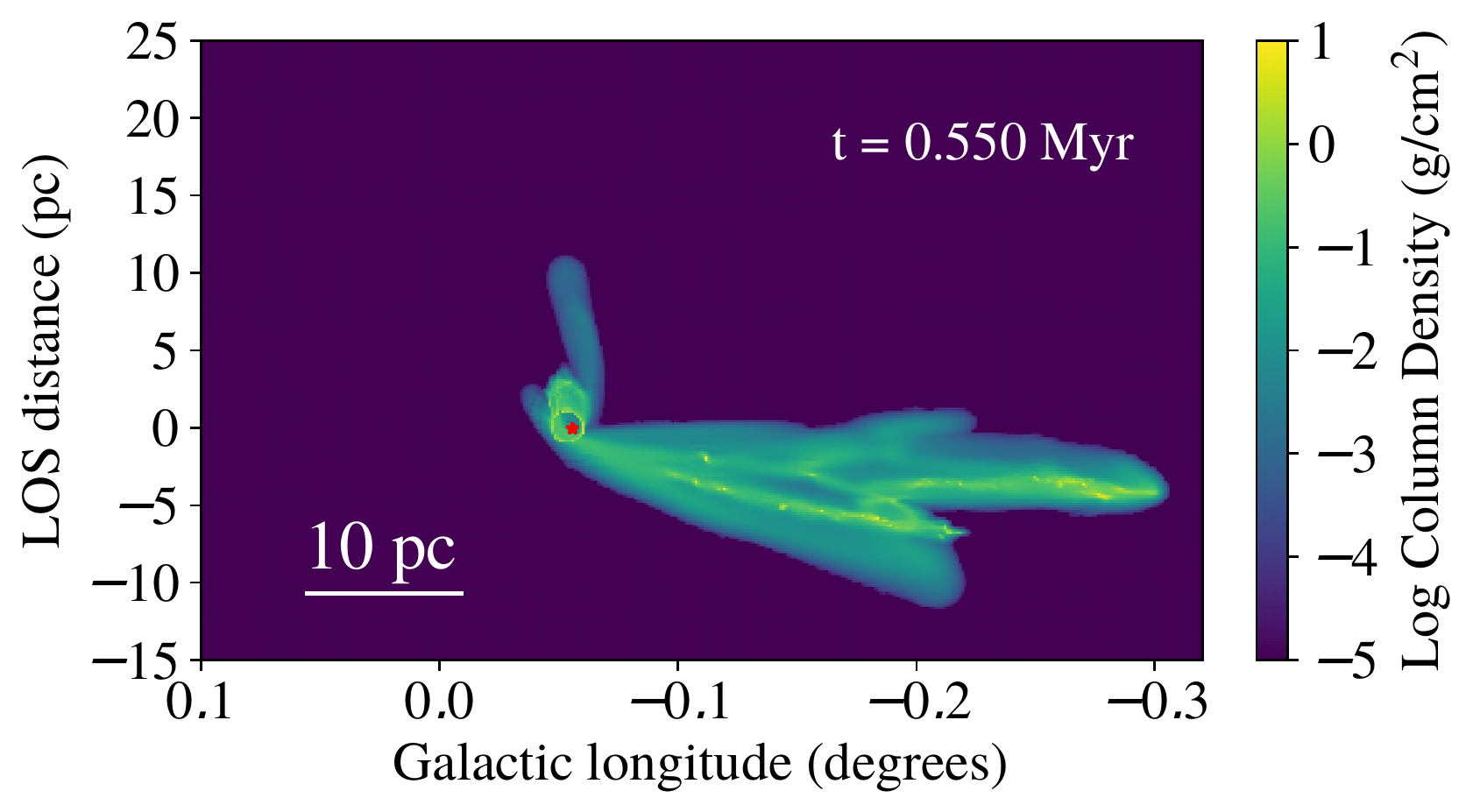}
\caption{Density projection along the Galactic latitude in the new reference system (see text in sections \ref{geomno50}) for our scenario. The axes represent the Galactic longitude (in degrees) and the line-of-sight distance from SgrA* (in pc). The red star labels the position of the SMBH.
}\label{models_above}
\end{center}
\end{figure}

\section{Discussion}\label{compar}

We have proposed a scenario to explain the simultaneous formation of the CND and the +20 km/s cloud via the tidal disruption of a large molecular cloud. In this scenario, the +20 km/s cloud reaches the Galactic centre from the south-east of the CND \citep[as already suggested by several other authors; e.g.,][]{Okumura91,Coil99}.

In particular, our ``picture'' is somewhat similar to one of the models presented in \citet{Vollmer02}. However, in our simulation the CND is directly formed by the plunge of the cloud, while in \citet{Vollmer02} it is assumed to be formed at earlier times.

First of all, as a general remark, we must stress again that the initial conditions of our simulation were not fine-tuned, hence we do not expect every observational feature to be perfectly matched. For example, the total mass of the +20 km/s cloud and the CND is estimated to be around $4\times 10^5 M_{\odot}$ (\citealp{Ferriere12}; however, this values are controversial; see, e.g., \citealp{Tsuboi18}), while our simulated cloud has an initial mass of only $10^5 M_{\odot}$. Nonetheless, our scenario depends mostly on the orbit of the cloud, hence we do not expect a significant change of our results under the assumption of an initially higher cloud mass. 

We must also acknowledge that, compared to other models, in this scenario the +20 km/s cloud reaches SgrA* with a much closer pericentre. In fact, most of the currently available models connect the +20 km/s and the +50 km/s cloud on the south-east of SgrA*, through the so-called molecular ridge. This hypothesis is indeed supported by the presence of the molecular ridge and by velocity maps showing a smooth velocity gradient between the two clouds \citep[for the most recent velocity maps, see, e.g.,][]{Liu12, Hsieh17}. On the other hand, the presence of the southern streamer \citep[e.g.,][]{Okumura91,Coil99} and of several smaller streamers (such as W-1, W-2, W-3 and W-4 shown in Fig. 2 of \citealp{Liu12} and Fig. 4 of \citealp{Hsieh17}) suggest a much ``closer'' interaction between the +20 km/s cloud and the CND, which is naturally provided in our scenario.

A connection of the CND and the +20 km/s to the +50 km/s in the framework of a single cloud plunge is possible. In fact, as shown in Fig. \ref{proj_with50}, when passing pericenter the cloud might appear as two different clouds, connected by a streamer of gas. Fig. \ref{proj_with50} might visually remind of the actual projected positions of the CND and the two observed molecular clouds. However, reconciling this configuration with the observations is more problematic (though we try to provide a more through comparison and a discussion of pros and cons of this further possible scenario in Appendix \ref{app50}).

A further constraint from observations could come from the estimated line-of-sight distance from SgrA*. Unfortunately, there is no agreement on the exact position of the +20 km/s cloud along the line-of-sight. For example, \citet{Vollmer03} put the +20 km/s in front of SgrA*, at a maximum line-of-sight distance of [-50,-25] pc. On the other hand, \citet{Ferriere12}, based on a collection of studies from different authors, prefers to place the +20 km/s cloud still in front of SgrA*, but at a distance of its center of [-12,-4] pc. In Fig. \ref{models_above}, we show a density map of our scenario, projected along the Galactic latitude. As visible, the +20 km/s cloud lies in front of SgrA*, at distances that are compatible with the estimates of \citet{Ferriere12}.

Finally, we must acknowledge that our scenario does not reproduce all the observed features \citep[e.g., it does not reproduce the NLE found by][]{Takekawa17}. This is possibly due to some simplified treatment of the involved physics. Nonetheless our model is quite succesful in simultaneously explaining many of them. Furthermore, our conclusions could also be affected by the supernova remnant SgrA East. This supernova remnant has an estimated age of $\gtrsim 10^4$ yr, which is much shorter than the orbital time of our cloud. As already shown by, e.g., \citet{Coil00} and \citet{Herrnstein05}, some of the observed spatial and velocity features, close to the CND, might also be explained by the interaction between SgrA East and the observed molecular complex. The inclusion of such supernova explosion in our simulation might be part of a follow-up study.

\section{Summary}

Our simulation of a molecular cloud infalling towards SgrA* matches many features of the observed CND and +20 km/s cloud. In particular, it can

\begin{itemize}
\item explain the formation of the CND;
\item link the CND to the + 20 km/s cloud, as suggested, e.g., by \citet{Takekawa17};
\item show that star formation in the +20 km/s cloud is possible, as indicated by the recent work by \citet{Lu15}. 
\end{itemize}

Specifically, the + 20 km/s cloud and the CND might both be the result of a single tidal disruption event of a large molecular cloud. The present work is to be thought as a quite effective proof of concept to explain such scenario.

More sophisticated simulations should include additional complex physical ingredients, such as a more realistic equation of state, a complex chemical network to study different molecular species separately and their thermodynamical impact on the gas and the radiative and mechanical feedback from stars. The latter ingredient, in the form of supernova explosion, would be, for example, crucial to try to account for the effect of SgrA East on the molecular gas complex. 
Nonetheless, the amount of observed features explained by the present model is quite remarkable, considering these limitations and that our initial conditions were not fine-tuned.

\section*{Acknowledgements}

AB and MM acknowledge financial support from the MERAC Foundation, through grant `The physics of gas and protoplanetary discs in the Galactic centre'. AAT acknowledges support from the JSPS KAKENHI Grant Number 17F17764. We would like to thank Prof. Romain Teyssier for his help with RAMSES and the whole ForDyS group for useful discussions. Most of the simulation post-processing was carried out with the yt toolkit \citep{Turk11}.




\bibliographystyle{mnras}
\bibliography{mylit} 




\appendix

\section{A further, temptative connection with the +50 km/s cloud}\label{app50}

As already discussed in section \ref{compar} and shown in Fig. \ref{proj_with50}, when the cloud is passing pericentre it may appear as two distinct clouds, connected close to the CND. In particular, Fig. \ref{proj_with50} reminds, in projection, the actual positions and sizes of both the +20 km/s and +50 km/s clouds \citep[cf. with][]{Novak00,Ferriere12,Mapelli16b}. This plot was obtained from the same simulation used for our main scenario, but choosing a different projection. In this case, we had to choose $(x,y,z)_{\hat{s}}\approx(0.400,-0.343,-0.850)$, $(x,y,z)_{\hat{l}}\approx(-0.404,-0.898,0.172)$ and $(x,y,z)_{\hat{b}}\approx(0.823,-0.274,0.498)$. In this different projection, the three orbital orientation angles (see section \ref{results}) are $i\approx 58.3^{\circ}$, $\Omega\approx 253.1^{\circ}$ and $\omega\approx 36.5^{\circ}$. 

Such alternative configuration has several interesting features, but also shows inconsistencies with the properties of the observed molecular complex. In this Appendix we will discuss them and we will propose few speculations to possibly reconcile this more complex scenario with the observations.

For example, in this case the +20 km/s cloud has its closest approach to the SMBH in the north of the CND. Recent observations might show indications of this being the case (cf. with the bridge in \citealp{Takekawa17} and the extW streamer in \citealp{Hsieh17}). On the other hand, as already discussed in section \ref{compar}, most of the current models connect the +20 km/s and +50 km/s clouds on the south-east. In principle, this would also be possible in a single plunge scenario. However, compared to the configuration shown in Fig. \ref{proj_with50}, a different projection should be adopted and the infalling cloud should have a more circular orbit, with a much larger pericentre distance (comparable to the distance of the observed molecular ridge), in order to undergo a southern passage. In this case, it would be harder to explain the southern streamer and the formation of the CND with the plunge of these clouds. Another possibility is that the observed velocity pattern of the two molecular clouds (which seem to be connected by a smooth velocity gradient) is overlapping and confused with the overall motion of the central molecular zone at larger distances \citep[see, e.g.,][]{Henshaw16}.

\begin{figure}
\begin{center}
\hspace*{0.105cm}
\includegraphics[scale=0.55]{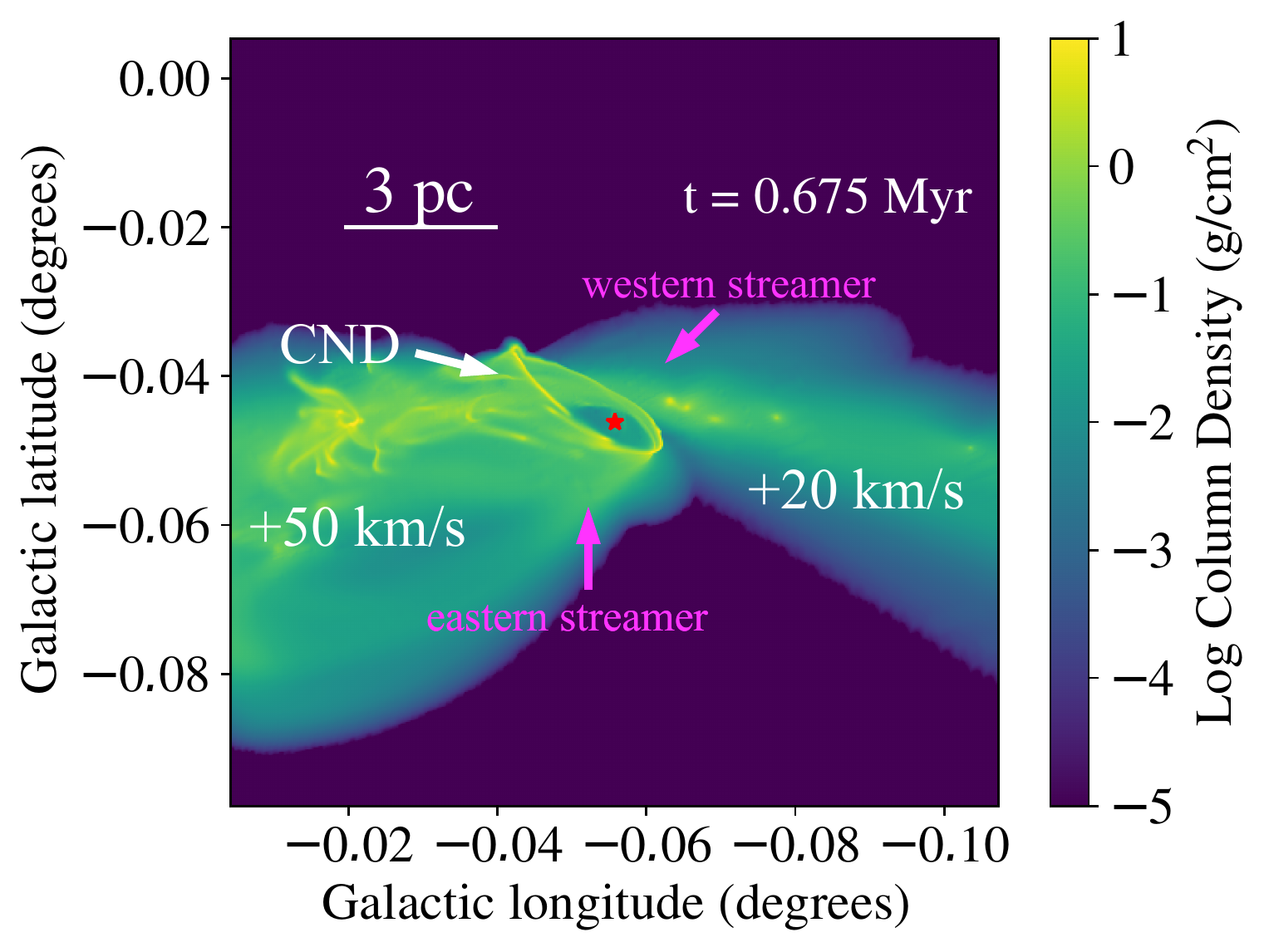}
\caption{Zoom-in of Fig. \ref{proj_with50}, around the inner 16 pc of the domain. The red star labels the position of the SMBH.
}\label{zoom_with50}
\end{center}
\end{figure}

As also visible in the zoom-in in Fig. \ref{zoom_with50}, in this different configuration, a stream of gas also appears to connect the +50 km/s and the CND. This might, in projection, be consistent with the so-called ``northern ridge'' \citep[e.g.,][and references therein]{McGary01,Ferriere12}, even though it is not matching its observed kinematical properties. In fact, this gas in the simulation has high (> +50 km/s) line-of-sight velocities, while the observed northern ridge has a bulk velocity of roughly -10 km/s, even though it seems to connect to the CND with a velocity gradient reaching +60 km/s \citep{McGary01}. We must stress that the region between the CND and the +50 km/s cloud is believed to be strongly affected by the supernova remnant SgrA East (see section \ref{intro}), so it is also difficult (and maybe deceptive) to strictly compare it to our simulation, which does not include such supernova explosion.

\begin{figure}
\begin{center}
\includegraphics[scale=0.55]{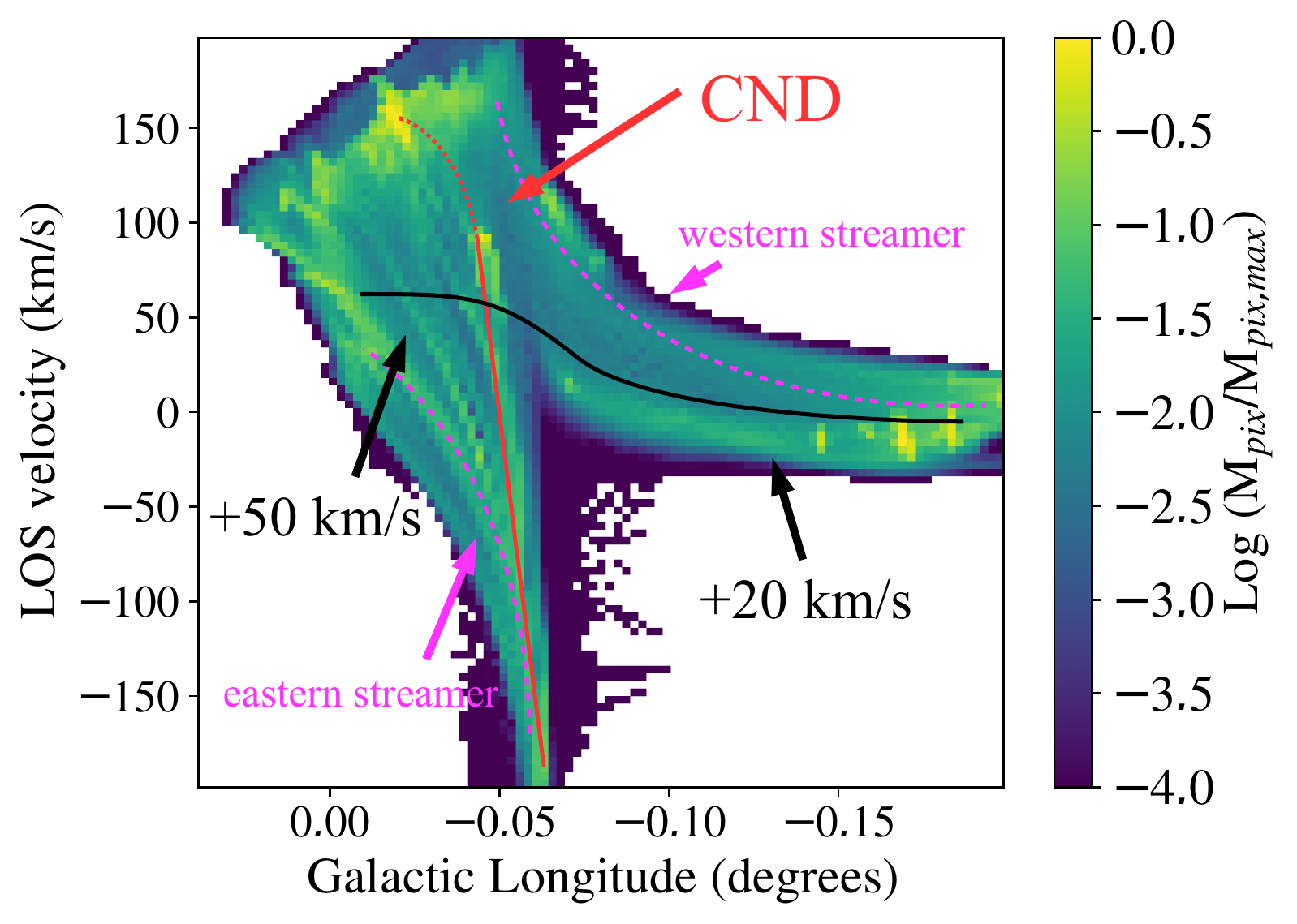}
\caption{Longitude-velocity diagram for the configuration in Fig. \ref{proj_with50}. The colormap shows the mass in every pixel of the diagram, scaled to the maximum value. At this time our simulation is able to reproduce some of the features of the observed CND, +20 km/s cloud and +50 km/s complex.
}\label{pv_with50}
\end{center}
\end{figure}

Fig. \ref{pv_with50} shows a longitude-velocity diagram for this different configuration. The black line shows the connection between the pre-pericentre and the post-pericentre parts of the cloud. Due to the fast orbital precession, the material that has passed pericentre keeps on moving with positive line-of-sight velocity. The post-pericentre gas has a bulk velocity that is a bit higher than +50 km/s and a large spread. A lot of gas is actually at velocities of around 150 km/s, which is much more than observed. Interestingly enough, this longitude-velocity diagram also shows motion of gas that is not fully coherent with the motion of the cloud and the CND (see the magenta dashed lines). These might be compared with the longitude-velocity diagrams of the extW and eastern streamers in \citet{Hsieh17} \citep[Fig. 5, 6 and 8 of their paper; see also the longitude-velocity diagrams in Fig. 2 of][]{Oka11}. Hence, in this more complex configuration, the extW streamer could be part of the +20 km/s cloud reaching the CND from ``above'', while \citep[as in the model of][]{Hsieh17} the eastern streamer could be material of the +50 km/s cloud, falling back towards the CND.

Nonetheless, the major problem of this simulated diagram is the too high velocity observed on the side of the +50 km/s cloud. This might be due to drag of part of the CND (see the dotted branch of the red line in Fig. \ref{pv_with50}). We must remind the reader that this is a line-of-sight velocity, thus it also depends on the actual trajectory followed by the cloud. The two observed clouds have relatively small line-of-sight velocities, despite their distance from the Galactic Centre. So, for such an eccentric orbit, lower line-of-sight velocities for the +50 km/s cloud could be obtained if the cloud trajectory lay closer to the plane of the sky. A faster precessing orbit could bring the second apocentre position away from the line-of-sight. This might be obtained by a higher stellar mass enclosed by the orbit, compared to our assumptions. The current precession angle is roughly 120$^{\circ}$ and, on a 1st order approximation, the precession angle scales linearly with the enclosed mass \citep[see, e.g.,][]{Gualandris12}. Hence, the enclosed stellar mass should be higher within a factor $\approx 1.5$, which, nonetheless, seems to be within the uncertainties of the currently available observations \citep{Chatzopoulos15, Fritz16, FeldmeierKrause17, GallegoCano18}.

\begin{figure}
\begin{center}
\includegraphics[scale=0.5]{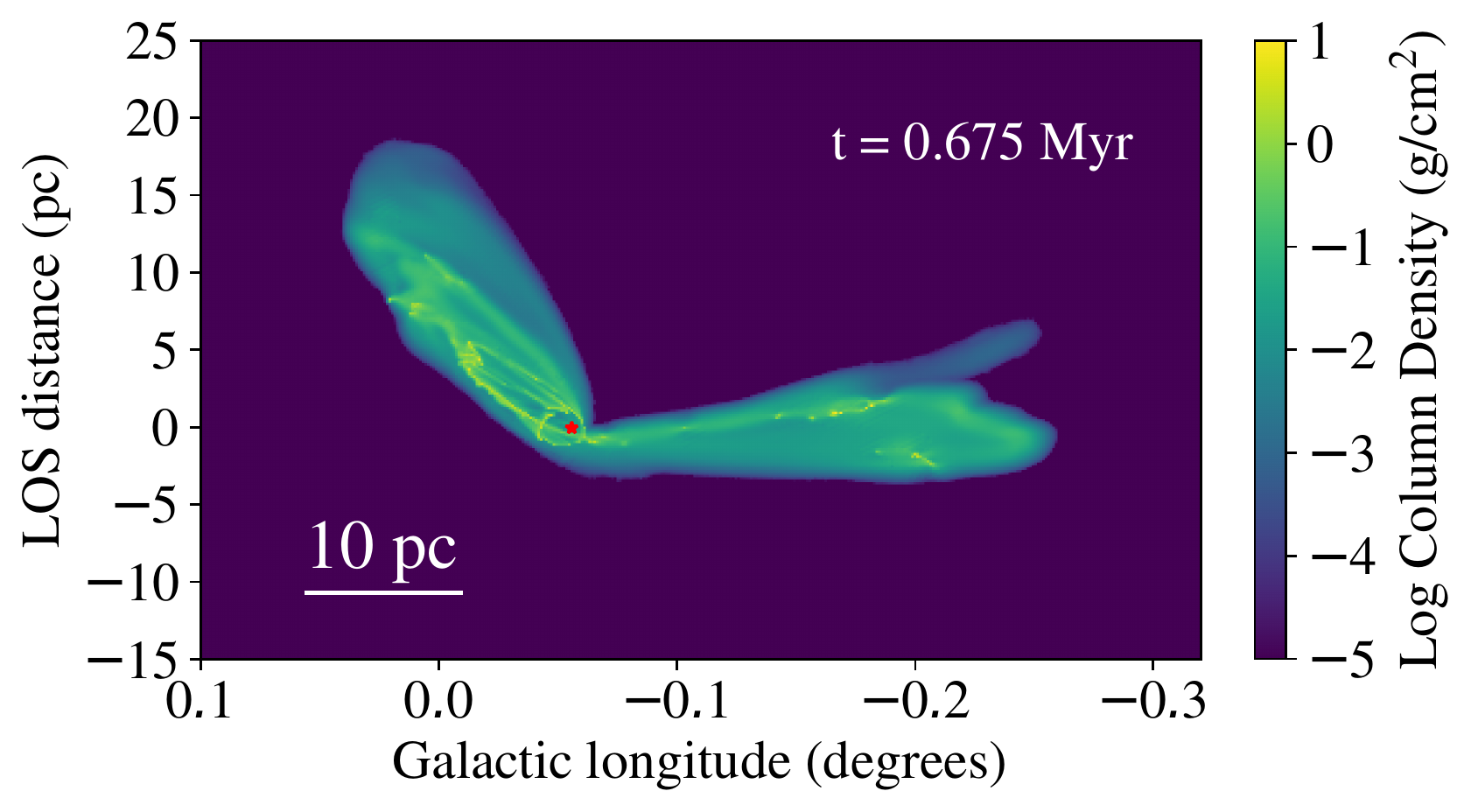}
\caption{Density projection along the Galactic latitude in the new reference system for the configuration in Fig. \ref{proj_with50}. The axes represent the Galactic longitude (in degrees) and the line-of-sight distance from SgrA* (in pc). The red star labels the position of the SMBH.
}\label{with50_above}
\end{center}
\end{figure}

In terms of line-of-sight distance from SgrA*, \citet{Vollmer03} put the +50 km/s cloud in front of SgrA*, at a maximum line-of-sight distance of -5 pc, while \citet{Ferriere12} prefers to place the +50 km/s cloud behind SgrA*, at a distance of 3 pc. As visible in Fig. \ref{with50_above}, the +20 km/s cloud lies roughly at the same distance from SgrA* as for our main scenario (cfr. Fig. \ref{models_above}), while the +50 km/s lies behind it, but it extends along the line-of-sight up to 15-20 pc from SgrA*, which is in disagreement with both the aforementioned estimates. Again, a faster precessing orbit might bring the post-pericentre trajectory of the infalling cloud closer to the plane of the sky, eventually giving the +50 km/s cloud a smaller line-of-sight extent and distance from SgrA*.

\bsp	
\label{lastpage}
\end{document}